\documentclass[12pt,preprint]{aastex}
\begin{document}
\tighten
\title{Magnetic Fields in Stellar Jets}

\author{
    Patrick Hartigan \altaffilmark{1},
    Adam Frank\altaffilmark{2},
        Peggy Varni\'ere\altaffilmark{2},
    and Eric G. Blackman\altaffilmark{2}
    }

\vspace{1.0cm}

\altaffiltext{1}{Dept. of Physics and Astronomy, Rice University,
6100 S. Main, Houston, TX 77005-1892, USA}

\altaffiltext{2}{Dept. of Physics and Astronomy, University of Rochester,
Rochester, NY 14627-0171}

\begin{abstract}

Although several lines of evidence suggest that jets from young stars are
driven magnetically from accretion disks, existing observations of field strengths
in the bow shocks of these flows imply that magnetic fields play only a minor role
in the dynamics at these locations. To investigate this apparent discrepancy we
performed numerical simulations of expanding magnetized jets with stochastically variable
input velocities with the AstroBEAR MHD code.  Because the magnetic field B is
proportional to the density n within compression and rarefaction regions,
the magnetic signal speed drops in rarefactions and increases in the compressed
areas of velocity-variable flows.  In contrast, B $\sim$ n$^{0.5}$ 
for a steady-state conical flow with a toroidal field, so the Alfven speed in
that case is constant along the entire jet. The simulations show that the combined
effects of shocks, rarefactions, and divergent flow cause magnetic fields
to scale with density as an intermediate power 1 $>$ p $>$ 0.5.  Because
p $>$ 0.5, the Alfven speed in rarefactions decreases on average as the jet
propagates away from the star.  This behavior is extremely important
to the flow dynamics because it means that a typical Alfven velocity
in the jet close to the star is significantly larger than it is in the rarefactions
ahead of bow shocks at larger distances, the one place where the field is
a measurable quantity. Combining observations of the field in bow shocks with 
a scaling law B $\sim$ n$^{0.85}$ allows us to infer field strengths
close to the disk. We find that the observed values of weak fields
at large distances are consistent with strong fields required to drive
the observed mass loss close to the star.  The increase of magnetic 
signal speed close to the star also means that typical velocity perturbations
which form shocks at large distances will produce only magnetic waves
close to the star.  For a typical stellar jet the crossover point inside which
velocity perturbations of 30 $-$ 40 km$\,$s$^{-1}$ no longer produce shocks is
$\sim$ 300~AU from the source.

\end{abstract}


\section{Introduction}

Emission line images of star forming regions often reveal spectacular collimated,
supersonic jets that emerge along the rotation axes of protostellar accretion disks
\citep[see][for reviews]{rb01,ray06}.
The jets break up into knots which form multiple bow shocks as faster
material overtakes slower material \citep[e.g.][]{hartigan01}.
Although measurements are scarce, when detected magnetic fields ahead of bow shocks
are weak; hence, the dynamics of the bow shocks are controlled by velocity perturbations
rather than by any magnetic instabilities.  In these systems the magnetic field
affects the flow mainly by reducing the compression in the dense postshock regions by adding
magnetic pressure support \citep{morse92,morse93}.

However, close to the star there is evidence that magnetic fields may dominate
the dynamics of jets.  Strong observational correlations
exist between accretion and outflow signatures \citep{cabrit90,heg},
and most mechanisms for accelerating jets from disks
involve magnetic fields \citep{ouyed97a,casse00}. 
Recent evidence for rotation in jets \citep{bacc02,coffey04} suggests
that fields play an important role in jet dynamics, at least
in the region where the disk accelerates the flow.

There has been considerable work done on the propagation of radiative jets
with strong ($\beta$ $\lesssim$ 1) magnetic fields \citep{cerq98,frank98,frank99,frank00,
gardiner00,gf00,or00,stone00,cerq99,cerq01a,cerq01b,colle06}.
These studies have tended to explore how magnetic fields influence the
large scale structure of jets, with the hope that 
the shape of jets may constrain the strength of the
magnetic fields.  These papers explored different field geometries, including
ones connected to magneto-centrifugal launch models.  Early studies
focused on the development of nose-cones, which form when toroidal magnetic
field is trapped due to pinch forces at the head of the flow.
The role of toroidal fields acting as shock absorbers within
internal working surfaces has also been explored by a number of authors.
More recent studies have focused on the H$\alpha$ emission
properties of MHD jets.

These papers did not, however, address the principle question of the
current work, which is to link together measurements of the field strengths
at different locations in real YSO jets and to infer the
global run of the magnetic field and density with distance from the source.  While
earlier studies \citep{gf00,or00,cerq01a} did explicitly identify the crucial connection
between internal working surfaces and magnetic field geometry when the
initial field is helical, the effect this would have on the dependence of
B($\rho$) and hence B(r) in a velocity-variable flow
was not considered, nor was the possibility of a `magnetic
zone' close to the source where $v_{shock}\ \sim\ V_A$. The realization that
such a region may have dynamically differentiable properties from the
super-fast zones downstream is, to the best of our knowledge, new to this
paper.  Thus, the work we present here represents the first attempt to
consider how the sparse magnetic fields measurements available in real YSO
jets can be used to infer large scale field patterns in these objects.

In what follows we show that magnetically dominated outflows
close to the disk are consistent with observations of
hydrodynamically dominated jets at larger distances,
provided the jets vary strongly enough in velocity to generate 
strong compressions and rarefactions.  
We begin by summarizing typical parameters of stellar jets, and then consider
what these numbers imply for the MHD behavior of a jet as a function of its distance
from the source for both the steady-state and time variable cases.

\section{Observed Parameters of Stellar Jets}

\subsection{Velocity Perturbations}

Stellar jets become visible as material passes through shock waves
and radiates emission lines as it cools. Flow velocities, determined
from Doppler motions and proper motions, are typically $\sim$ 300
km$\,$s$^{-1}$. The emission lines are characteristic of much
lower shock velocities, $\sim$ 30 km$\,$s$^{-1}$ in most cases,
leading to the idea that small velocity perturbations on the order
of 10\%\ of the flow speed (with occasional larger amplitudes as high
as 50\%)
continually heat the jet \citep{rb01}.

For jets like HH~111 which lie in the plane of the sky
we can observe how the velocity varies at each point along the
flow in real time by measuring proper motions of the emission.
Thanks to the excellent spatial resolution of the Hubble Space Telescope,
errors in these proper motions measurements are now only $\sim$ 5 km$\,$s$^{-1}$,
which is low enough to discern real differences in the velocity of
material in the jet.  As predicted from
emission line studies, the observed differences between adjacent knots
of emission are typically 30 $-$ 40 km$\,$s$^{-1}$ \citep{hartigan01}.

\subsection{Density}

Opening angles of stellar jets are fairly constant along the
flow, ranging between a few degrees to $\gtrsim$ 20 degrees
\citep[e.g.][]{rb01,cdr04}. Hence, to a good approximation we can
take the flow to be conical.  Once the jet has entered a strong
working surface it splatters to the sides, making its width
appear larger, so the most reliable measures of jet widths
are those close to the source. Other effects,
such as precession of the jet and inhomogeneous ambient media
also influence jet widths at large distances.
In the absence of these effects, stellar jets can stay collimated for large
distances because they are cool $-$ the sound speeds of $\lesssim$ 10 km$\,$s$^{-1}$
are small compared with the flow speeds of several hundred km$\,$s$^{-1}$.

A well-known example of a conical flow is HH~34, which has a bright
jet that has a nearly constant opening angle until it reaches
a strong working surface \citep[cf. Figure 6 of ][]{reip02}.
If we extend the opening angle defined by the sides of the jet 
close to the source to large enough distances to meet the large bow shock HH~34S,
we find that the size of the jet at that distance
is close to that inferred for the Mach disk of
that working surface \citep{morse92}, as expected for a conical flow.

If jets emerge from a point then
the density should be proportional to r$^{-2}$ except perhaps
within a few AU of the source where the wind is accelerated.
New observations of jet widths range from a few AU at the source,
to as high as 15~AU for bright jets like HH~30. For a finite source
region of radius h, the density n $\sim$ $(r+r_0)^{-2}$ for a conical
flow, where $r_0$ = h/$\theta$, and $\theta$ is the half opening angle of the
jet. For h = 5~AU and $\theta$ = 5 degrees, $r_0$ = 57~AU.

For the purposes of constructing a set of fiducial values for jets,
we adopt a density of $10^4$ cm$^{-3}$ at 1000~AU, and assume the width of
the jet at the base to be 10 AU, with an opening half-angle of 5 degrees.
These parameters produce a mass loss rate of
$5\times 10^{-8}$ M$_\odot$yr$^{-1}$ for a flow velocity of 300 km$\,$s$^{-1}$.
With these values we can calculate densities as a function of distance
(the third column of Table~1).  The fiducial values in the Table are
only a rough guide to the densities observed in a typical jet.
In addition to intrinsic variations between objects and density variations
lateral to the jet, beyond $\sim$ 1000~AU the observed densities
increase substantially over a volume-averaged density in the Table owing to compression
in the cooling zones of the postshock gas. Densities are correspondingly lower
in the rarefaction regions between the shocks.

The density dependence in Table 1 for a conical flow appears about right from the
data. New observations of the electron densities and ionization fractions
at distances of $\sim$ 30~AU of the jet in HN~Tau indicate a total density
between $\sim$ 2$\times 10^6$ cm$^{-3}$ and $10^7$ cm$^{-3}$ \citep{hep04},
while the average density in jets such as HH~47, HH~111,
and HH~34 at $\sim$ $10^4$~AU are
$10^3$ -- $10^4$ cm$^{-3}$ \citep[Table 5 of][]{hmr94}.

\subsection{Magnetic Field}

Because most stellar jets radiate only nebular emission lines, which are unpolarized and
do not show any Zeeman splitting, measurements of magnetic fields in jets
are not possible except for a few special cases. The only measurement
of a field in a collimated flow close to the star appears to be that of \citet{ray97}, who
found strong circular polarization in radio continuum observations of T~Tau~S.
The left-handed and right-handed circularly polarized light appear offset
from one another some 10~AU on either side of the star, and the degree of
polarization suggests a field of several Gauss. \citet{ray97} argue that the fields
are too large to be attached to the star, and must come from compressed gas behind
a shock in an outflow.
However, \citet{loinard05} interpret the extended continuum emission from this
object in terms of reconnection events at the star-disk interface.
If the emission does arise in a jet, then
even taking into account compression, the fields must
be at least hundreds of mG in front of the shock to produce the observations.

One other technique has been successful in measuring magnetic fields in jets,
albeit at larger distances.
As gas cools by radiating behind a shock, the density, and hence the
component of the magnetic field parallel to the plane of the shock
(which is tied to the density by flux-freezing) increases
to maintain the postshock region in approximate pressure equilibrium. As a result,
the ratio of the magnetic pressure to the thermal pressure scales as T$^{-2}$
\citep{hartigan03}, so at some point in the cooling zone the magnetic pressure
must become comparable to the thermal pressure
even if the field was very weak in the preshock
gas. The difference between the electron densities
inferred from emission line ratios such as [S~II] 6716/6731 for a nonmagnetic
and weakly-magnetized shock can be as large as two orders of magnitude. Hence,
one can easily measure the component of the magnetic field in the plane of the
shock by simply observing the [S~II] line ratio, provided the preshock density
and the shock velocity are known from other data.

The total luminosity in an emission line constrains the preshock density well,
so the problem comes down to estimating the shock velocity. For most jets this
is a difficult task from line ratios alone because spectra from shocks with
large fields and high shock velocities resemble those from small fields and
low shock velocities \citep{hmr94}. The easiest way to break this
degeneracy is if the shock is shaped like a bow and the velocity is large enough
that there is [O~III] emission at the apex. Emission lines of [O~III] are
relatively independent of the field, and occur only when the shock velocity
exceeds about 90 km$\,$s$^{-1}$. Hence, by observing how far [O~III] extends
away from the apex of the bow, and observing the shape of the bow, one
can infer the shock velocity. Combining the shock velocity, the preshock
density and the observed density in the cooling zone gives the magnetic field.

Unfortunately, only a few bow shocks have high enough velocities to emit
[O~III], so only HH~34S and HH~111V have measured fields. The two cases yield
remarkably similar results. In HH~34S, located
$5.1\times 10^4$~AU from the source, the preshock
gas has a density of 65 cm$^{-3}$ and a magnetic field of 10$\mu$G \citep{morse92},
while HH~111V is $6.4\times 10^4$~AU from the star and has a preshock density of
200 cm$^{-3}$ and a magnetic field of 30$\mu$G \citep{morse93}.

The ratio of B/n is the same for both HH~34S and HH~111V -- we
take 15$\mu$G at a density of 100 cm$^{-3}$ as a typical value.
To fill in the field strengths throughout the table requires a relationship
between B and n, which we now explore.

\section{The Scaling Law B $\sim$ n$^p$}

There are two analytical scaling laws between the magnetic field and the
density that might apply to stellar jets. If jets are driven by some sort
of disk wind, then at distances beyond the Alfven radius
\citep[typically a few AU,][]{anderson05}, the field will be mostly toroidal,
and should decline as r$^{-1}$ along the axis of the jet, where r is the
distance from a point in the jet to the source.
This radial dependence can be visualized
by taking a narrow slice of thickness dz perpendicular to the axis of the jet.
As the slice moves down the jet, its thickness remains constant because the
jet velocity is constant at large distances from the disk, and the diameter
of the slice increases linearly with the distance from the source as the
flow moves. Hence the cross sectional area of the slice increases linearly
with distance. The toroidal field strength, proportional to the number of
field lines per unit area in the slice, must therefore scale as r$^{-1}$.
A similar argument shows that the radial B scales as r$^{-2}$ for a conical
flow, which is why the toroidal field dominates in the jet outside of the
region near the disk. For a conical
flow, the density drops as r$^{-2}$, so B $\sim$ n$^{0.5}$ for a steady flow.
In contrast, if shocks and rarefactions dominate the dynamics, then
the field is tied to the density, so B $\sim$ n.

To determine which of these dependencies dominates we
simulated an expanding magnetized flow that produces shock waves from
velocity variability.  Our simulations are carried out in 2.5D using
the AstroBEAR adaptive mesh refinement (AMR) code. AMR allows high
resolution to be achieved only in those regions which require it due
to the presence of steep gradients in critical quantities such as gas
density. AstroBEAR has been well-tested on variety of problems in 1,
2, 2.5D \citep{var06} and 3D \citep{lebedev04}.  Here we use the MHD
version of the code in cylindrical symmetry (R,z) with {\bf B} =
B$_\phi${\bf e}$_\phi$, hence maintenance of $\nabla \cdot {\bf B} = 0$
is automatically achieved.  We initialize our jet with magnetic
field and gas pressure profiles ($B_\phi(R), P(R)$) which maintain
cylindrical force equilibrium \citep{frank98}.

The spatial scale of the grid is arbitrary, but for plotting purposes we
take it to be 10~AU so that the extent of the simulation resembles that of
a typical stellar jet. Choosing a scale of 1~AU would match the
dimensions at the base of the flow. The time steps are set to be 0.5 of the
Courant-Friedrich-Levy condition, which is the smallest travel time for
information across a cell in the simulation. For a 200 km$\,$s$^{-1}$ jet
and a 10~AU cell size this time interval is $\Delta$t = 0.12 years.
The input jet velocity is a series
of steps, whose velocity in km$\,$s$^{-1}$ is given by V = 200(1+fr), where f
is the maximum amplitude of the velocity perturbation,
and r is a random number between $-$1 and 1. We  
ran simulations with f = 0.5, 0.25, and 0.10.
We verified that a constant velocity jet gave a constant Alfven velocity and
n $\sim$ $(r+r_0)^{-2}$ as predicted by analytical theory.
The opening half angle of the jet was 5 degrees;
a numerical run with a wider opening half angle of 15 degrees produced the same
qualitative behavior as the more collimated models.

The first ten cells, taken to be the smallest AMR grid size, are 
kept at a fixed velocity V for the entire length of the pulse, and these
ten cells are overwritten with a new random velocity after a pulse
time of $\sim$ 7.2 years (60$\Delta$t) for a grid size of 10~AU and a
velocity of 200 km$\,$s$^{-1}$.  Densities, velocities, and magnetic field strengths
are mapped to a uniform spatial grid and
printed out whenever the input velocity changes.
Cooling is taken into account in an approximate manner by using a polytropic 
equation of state with index $\gamma$ = 1.1.
The density of the ambient medium is 1000 cm$^{-3}$ and the
initial density of the jet is held constant at 7500 cm$^{-3}$.
We fixed the initial magnetic field to give
a constant initial Alfven speed of 35 km$\,$s$^{-1}$. 

Figs.~1 $-$ 4 show the results obtained for the f = 0.5 case.
Similar plots were made for a single, nonmagnetic velocity perturbation in 1-D
by \citet{hr93}.  Positive velocity perturbations
form compression waves that steepen to form forward and reverse shocks (a bow
shock and Mach disk in 2D), while negative velocity perturbations produce
rarefactions as fast material runs ahead of slower material.
The top panel in Fig.~1 shows the density along the axis of the jet once
the leading bow shock has progressed off the grid.
The strongest rarefactions, marked as open
squares, follow closely to an r$^{-2}$ law.  Essentially once these
strong rarefactions form in the flow, the gas there expands freely until it
is overrun by a shock wave. Because each of the input velocity perturbations
begins by forcing a velocity into the first 10 AMR zones
(a region $\sim$ 100 AU from the source depending on the size of the AMR zone),
rarefactions caused by drops in the random velocity
originate from log(r) $\sim$ 2 (Fig.~1). Hence, the open squares
lie close to a line that goes through the steady-state solution at this point.

The bottom plot shows that shock waves and rarefactions dominate the flow
dynamics.  By the end of the simulation, the $\sim$ 35 perturbations have
interacted with one another, colliding and merging to 
create only seven clear rarefactions and a
similar number of shocks. The jet evolves quite differently than
it would in steady state (V$_A$ = constant). While the gas initially
follows a B $\sim$ n$^p$ law with p = 0.5, as soon as
shocks and rarefactions begin to form, the
value of p becomes closer to unity, with p $\sim$ 0.85 a reasonable match
to the entire simulation. 

The important point is that
once shock waves and rarefactions form, they will increase the value
of p above that expected for a steady state flow. This increase means that the
magnetic signal speed (a term that refers to fast magnetosonic waves,
slow magnetosonic waves, or Alfven waves, all of which have
similar velocities because the sound speed is low, $\sim$ 10 km$\,$s$^{-1}$)
drops overall at larger distances, especially within the
rarefaction waves. Hence, small velocity perturbations that form only magnetic
waves close to the star will generate shocks if they overrun rarefacted gas at
large distances from the star. Essentially velocity perturbations redistribute
the magnetic flux and thereby facilitate shock formation over much of the jet.

Using the numerical values from section 2.3, we can fill in the
fourth column in Table~1 using B/(15~$\mu$G) = (n/100 cm$^{-3}$)$^{0.85}$.
The fifth column of the Table gives the Alfven
speed in the preshock gas assuming full ionization, which is also appropriate for
dynamics of partially ionized gas as discussed below.

\section{Discussion}

\subsection{Evolution of a Typical Velocity Perturbation in an MHD Jet}

Following how individual velocity perturbations evolve with time illustrates many
of the dynamical processes that govern these flows. Fig.~2 shows a typical
sequence of such perturbations, labeled A, B, C, D, and E, with initial velocities
of 192, 230, 172, 223, and 295 km$\,$s$^{-1}$, respectively. 
In the left panel, which shows the simulation after 11 velocity pulses,
a compression zone (marked as a solid vertical line) forms as B overtakes A,
and both the density and Alfven velocity V$_A$ increase at this interface.
Other compression zones grow from the interfaces of E/D and D/C. 
The rarefaction (dashed line) between B and C creates a characteristic `ramp' profile
in velocity, and at the center of this feature lies a broad, deep density
trough an order of magnitude lower than the surrounding flow. The Alfven
speed in this trough has already dropped to nearly 10 km$\,$s$^{-1}$.
For comparison, the steady state solution has V$_A$ = 35 km$\,$s$^{-1}$
everywhere, with a density that declines from the input value of 7500 cm$^{-3}$.

The right panel shows the simulation several hundred time steps later,
after 12 velocity pulses have passed through the input nozzle of the jet.
Pulses A, B and C, have all evolved into something other than a step function,
and little remains of pulse D, which will soon form the site of a merger
between the denser knots at the D/E and C/D interfaces. The compression wave
between A and B (1125 AU at left, and 1475 AU at right) has an interesting
kink in its velocity profile. The two steep sides of this kink would become 
forward and reverse shocks if it were not for the fact that the
Alfven speed there remains high enough, $\sim$ 35 km$\,$s$^{-1}$, to
inhibit the formation of a shock. 

The left panel of Fig.~3 shows the same region of the jet several pulse times
later. The only remaining pulse in this section of the jet is E, which has formed both a forward
(bow) shock and a reverse (Mach disk) shock. The Alfven speed at $\sim$ 2300 AU
ahead of the forward shock and at $\sim$ 2100 AU behind
the reverse shock are both only 10 $-$ 20 km$\,$s$^{-1}$, so this gas
shocks easily. Both the forward and reverse shocks have magnetosonic
Mach numbers of 2 $-$ 3. The working surface between these shocks has a density
of $\sim$ $3\times 10^4$ cm$^{-3}$, a factor of 4 increase over the
initial jet density at the source and about two orders of magnitude higher
than the surrounding gas. The Alfven speed there is 120 km$\,$s$^{-1}$, 
having reached a maximum of 140 km$\,$s$^{-1}$ when the shock
first formed. Pulses A through C have merged to create a zone of nearly
constant velocity from 2400 $-$ 3400~AU. The density in this region is
far from constant, however, with the density in the feature at 2900 AU
a factor of 500 higher than its surroundings. This type of feature can
cause problems in estimating mass loss rates, because it is a dense
blob with substantial mass that is no longer being heated by shocks, and
may therefore not appear in emission line images.

The right panel of Fig.~3 shows the working surface of knot E after 3
more pulse times. The velocity perturbation E has weakened to $\sim$ 30 km$\,$s$^{-1}$
but still forms a pair of shocks because the surrounding gas has an Alfven speed of
only 10 km$\,$s$^{-1}$. The magnetic pressure in the working surface is
high enough to cause the region to expand, which lowers the density and
the Alfven speed. In the right panel the working surface is now 200 AU 
wide and the Alfven speed has dropped to about 70 km$\,$s$^{-1}$.
A new shock is just forming at 3900 AU as all the material on the
left side of the plot with V $>$ 200
km$\,$s$^{-1}$ overtakes slower, but relatively dense gas from 3900 AU to
4400 AU. 

The continuous creation and merging of shocks, rarefactions, and compression
waves leads to some interesting and unexpected results. Because dense knots can
have significant magnetic pressure support, when they collide they can `bounce',
as has been seen before in simulations of colliding magnetized clouds
\citep{miniati99}.  Evidence for splashback from such a collision
is evident later in the
simulation where the velocity at one point drops to 70 km$\,$s$^{-1}$,
lower than any of the input velocities, which all lie between 100 km$\,$s$^{-1}$
and 300 km$\,$s$^{-1}$.

Magnetically, the overall effect is to concentrate the field
into a few dense areas, which then subsequently expand \citep[see also][]{gf00}.
Fig.~4 shows the Alfven speed at end of the simulation, by which time the leading bow
shock has propagated off the right end of the grid. Though there are a few
areas that have large Alfven speeds, most of the gas in the jet has a
significantly lower V$_A$ than the steady-state solution does (solid line).
The graph shows that, on average, magnetic fields tend to be
more important dynamically close to the star.

Lower-amplitude simulations (Fig.~5) show similar qualitative behavior both in
the formation and propagation of shocks and rarefactions, and in 
the dependency of B \hbox{vs.} n. As expected, fewer shocks and rarefactions
form in the low-amplitude simulations and the results are closer to the 
steady-state solution (p = 0.5). In all cases, areas of high Alfven velocity concentrate
into a few shocked regions where the density is high, and most locations along
the jet have lower Alfven speeds than those of the steady state case.

\subsection{The Hydrodynamic and Magnetic Zones}

As noted in section 3 and in Figs.~4 and 5, because
B $\sim$ n$^p$ along the jet with $p>0.5$, the Alfven
speed V$_A$ $\sim$ increases, as the density rises. When n $\gtrsim$ $10^5$ cm$^{-3}$, a typical
velocity perturbation of 40 km$\,$s$^{-1}$ will produce a magnetosonic wave rather
than a shock.  This variation of the average magnetic signal speed with density, and therefore with
distance, implies that jets can behave hydrodynamically 
at large distances, and magnetically close to the star.

Far from the star, the densities are low and the
dynamics are dominated by multiple bow shocks and rarefactions that
form as faster material overtakes slower material. The
magnetic field reduces the compression in the cooling
zones behind the shocks and cushions any collisions between knots,
but is otherwise unimportant in the dynamics. 
The fiducial values in Table~1 show that this hydrodynamic
zone typically extends from infinity to within about 300~AU
of the star ($\sim$ 1$^{\prime\prime}$ for a typical source),
so most emission line images of jets show gas in this zone.
Alternatively, when the magnetic signal speed is greater than a typical
velocity perturbation, the magnetic field inhibits the formation of
a shock unless the perturbation is abnormally large.
Figs. 4 and 5 show that the boundary between the magnetic and hydrodynamic
zones is somewhat ill-defined:
magnetic forces dominate wherever the field is
high enough, as occurs in a few places in the simulations at
large distances, for example, in the aftermath of the collision
of two dense knots. However, statistically we expect magnetic fields
to prevent typical velocity perturbations from forming shocks
inside of $\sim$ 300~AU.

A potential complication with the above picture is that fields may dampen
small velocity perturbations in the magnetic zone before the
perturbations ever reach the hydrodynamic zone where they are able to create shocks.
How such perturbations behave depends to a large degree on how disk winds initially
generate velocity perturbations in response to variable disk accretion rates.
If the mass loss is highly clumpy, then plasmoids of dense magnetized gas may simply
decouple from one another at the outset, produce rarefactions, and thereby
reduce the magnetic signal speed enough to allow the first shocks to form. 
In addition, the geometry of the field will
not remain toroidal if the flow becomes turbulent owing
to fragmentation, precession, or interactions between clumps.
When both toroidal and poloidal fields are present, velocity variability
concentrates the toroidal fields into the dense shocked regions and the poloidal
field into the rarefactions \citep{gf00}. The magnetic signal speed in poloidally-dominated
regions drops as the jet expands, facilitating the formation of shocks in these
regions.

It might be possible to confirm the existence of stronger fields in knots
close to the source with existing instrumentation. As described in section 2.3,
by combining proper motion observations with emission line studies 
one can infer magnetic fields provided the velocity perturbations have
large enough amplitudes. 

\subsection{Connection to the Disk}

At distances closer to the disk than 10~AU, a conical flow with a finite
width (n $\sim$ $(r+r_0)^{-2}$)
is not likely to model the jet well. For a disk wind, the field
lines should curve inward until they intersect the disk at $\lesssim$ 1~AU,
while the field changes from being toroidal to mostly poloidal.
We can use the scaling law between magnetic field strength and density derived above
to see if the field strengths are roughly consistent with an MHD launching scenario.
With B $\sim$ n$^{0.85}$, the Alfven velocity equals the jet speed,
$\sim$ 300 km$\,$s$^{-1}$, when n $\sim$ $4\times 10^7$cm$^{-3}$ and
B $\sim$ 0.9~G. A moderately strong shock could then increase the field strength
to a few Gauss, as observed by
\citet{ray97}. Taking the density proportional to r$^{-2}$ within 10~AU
gives r = 2.5~AU when v = 300 km$\,$s$^{-1}$,
the correct order of magnitude for the Alfven radius of an MHD disk wind.
The footpoint of the field line in the disk would be $\sim$ 0.4~AU for a
central star of one solar mass.

The observed correlation of accretion and outflow signatures, together
with the existence of a few very strong bow shocks in some jets, suggests
that sudden increases in the mass accretion rate through the disk produce
episodes of high mass loss that form knots in jets as the material
moves away from the star. Young stars occasionally exhibit large accretion
events known as FU~Ori and EX~Ori outbursts \citep{hartmann04,briceno04}, that
may produce such knots.
However, because knots typically take tens of years to move far enough away
from the star to be spatially resolved, it has been difficult to tie an
accretion event to a specific knot in a jet. In the case of a newly-ejected
knot from the T Tauri star CW~Tau, there does not appear to have been
an accretion event at the time of ejection, though the photometric
records are incomplete \citep{hep04}.

Because magnetic fields must dominate jets close to the disk, it is
possible that the origin of jet knots is purely magnetic.
Models of time-dependent MHD jets have
produced knots that are purely magnetic in nature, and do not require
accretion events \citet{ouyed97b}.  For this
scenario to work the mechanism of creating the knots must also
impart velocity differences on the order of 10\%\ of the flow velocity
in order to be consistent with observations of velocity variability at
large distances from the star. It may also be necessary to
decouple the field from the gas via ambipolar diffusion in order to
reduce the Alfven speed enough to allow these velocity
perturbations to initiate shocks and rarefactions.  However, ambipolar
diffusion timescales appear to be too long to operate efficiently in jet beams
\citep{frank99}. One way
to distinguish between accretion-driven knots and pure
magnetic knots is to systematically monitor the brightness of T~Tauri
stars with bright forbidden lines over several decades to see whether
or not accretion events are associated with knot ejections.

\subsection{Effects of Partial Preionization}

The ionization fraction of a gas affects how it responds to magnetic disturbances.
Cooling zones of jets are mostly neutral --
the observed ionization fractions of
bright, dense jets range from $\sim$ 3\%\
-- 7\%\ \citep{hmr94,podio06}, and rise to $\sim$ 20\%\ for some objects \citep{be99}.
The ionization fraction is higher close to star in some jets, $\sim$
20\%\ if the emission comes from a shocked zone, and
as much as 50\%\ for a knot of uniform density \citep{hep04}, while in 
HH~30 the ionization fraction rises from a low value of $\lesssim$ 10\%\
to about 35\%\ before declining again at larger distances \citep{hm07}.

The Alfven speed in a partially ionized gas like a stellar jet
is inversely proportional to the density of ions,
not to the total density. If the Alfven speed exceeds the shock velocity,
then ions accelerated ahead of the shock collide with neutrals and form a
warm precursor there. If the precursor is strong enough
it can smooth out the discontinuity of the flow variables at the shock front
into a continuous rise of density and temperature known as a C-shock
\citep{draine80,draine83}. Precursors have been studied when the gas is molecular
\citep{flower03,ciolek04}, but we have not found any calculations of
the effects precursors have on emission lines from shocks when the
preshock gas is atomic and mostly neutral.

Dynamically the main issue is whether or not the magnetic signal speed in the preshock
gas is large enough to inhibit the formation of a shock. Because ions
couple to the neutrals in the precursor region via strong charge exchange
reactions, any magnetic waves in this region should be quickly mass-loaded with neutrals.
Hence, the relevant velocity for affecting the dynamics is the Alfven velocity
calculated from the total density, and not the density of the ionized
portion of the flow. Another way to look at the problem is to consider the
compression behind a magnetized shock, taking a large enough grid size so
the precursor region is unresolved spatially. By conserving mass, momentum,
and energy across the shock one finds that the compression in a magnetized
shock varies with the fast magnetosonic Mach number in almost an identical way
that the compression in a nonmagnetized shock varies with Mach number
\citep[Figure 1 of][]{hartigan03}. Hence,
the effective signal speed that determines the compression is
calculated using the total density, and not the density of the ionized component.
For this reason we use
the total density to calculate the Alfven speed in the fifth column of
Table~1.

\section{Summary}

We have used observations of magnetic fields and densities in stellar
jets at large distances from the star to infer densities and field
strengths at all distances under the assumptions of a constant opening
angle for the flow and flux-freezing of the field.  Numerical simulations
of variable MHD jets show that shocks and rarefactions dominate the
relation between the density n and the magnetic field B, with the relation
approximately B $\sim$ n$^p$, with 1 $>$ p $>$ 0.5. Because p $>$ 0.5, the
Alfven velocity increases at higher densities, which occur on average closer to
the star.  This picture of a magnetically dominated jet close to the star that gives
way to a weakly-magnetized flow at larger distances is consistent with
existing observations of stellar jets that span three orders of magnitude
in distance.  Velocity perturbations effectively sweep up the magnetic flux
into dense clumps, and the magnetic signal speed drops markedly in the
rarefaction zones between the clumps, which allows shock waves to form easily
there. For this reason, magnetic fields will have only modest dynamical effects on
the visible bow shocks in jets, even if fields are dynamically important 
in a magnetic zone near the star.  

\acknowledgements{This research was supported in part by a NASA grant from
the Origins of Solar Systems Program to Rice University. We thank Sean Matt
and Curt Michel for useful discussions on the nature of magnetic flows.
}

\null
\begin{center}
\begin{deluxetable}{ccccc}
\singlespace
\tablenum{1}
\tablewidth{6.5in}
\tablecolumns{5}
\tabcolsep = 0.06in
\parindent=0em
\tablecaption{}

\startdata
\noalign{\medskip}
\noalign{Average Jet Parameters}
\noalign{\medskip}
\noalign{\hrule}
\noalign{\medskip}
Distance From Star (AU) & Arcseconds$^a$ & n (cm$^{-3}$)$^b$ & B$_\perp$ & V$_A$ (km s$^{-1}$)$^c$\\
\noalign{\smallskip}
\noalign{\hrule}
\noalign{\smallskip}
10             &  0.02  &  $2.5\times 10^6$   & 82 mG         & 113  \\
30             &  0.06  &  $1.5\times 10^6$   & 53 mG         & 94  \\
100            &  0.2   &  $4.5\times 10^5$   & 19 mG          & 62 \\
300            &  0.6   &  $8.8\times 10^4$   & 4.8 mG          & 35 \\
$10^3$         &  2.2   &  $10^4$             & 0.75 mG         & 16 \\
$3\times 10^3$ &  6.5   &$1.2\times 10^3$$^d$ & 124 $\mu$G$^d$ & 7.8 \\
$10^4$         &  22    &    110$^d$          & 16$ \mu$G$^d$  & 3.3 \\
$3\times 10^4$ &  65    &     12$^d$          & 2.4$ \mu$G$^d$ & 1.5 \\
\enddata
\tablenotetext{a}{Spatial offset from the star at the distance of the Orion star forming region (460 pc).}
\tablenotetext{b}{Densities for a conical flow with a half
opening angle of 5 degrees and a base width of 10~AU,
taking the density to be $10^4$ cm$^{-3}$ at 1000~AU.}
\tablenotetext{c}{The Alfven speed V$_A$ determined from the total density n.}
\tablenotetext{d}{Values refer to an average density; densities at large distances
are highly influenced by shocks and rarefaction waves, see text.}
\end{deluxetable}
\end{center}
\vfill\eject

\begin{figure}

\vbox to 6.0in{\includegraphics{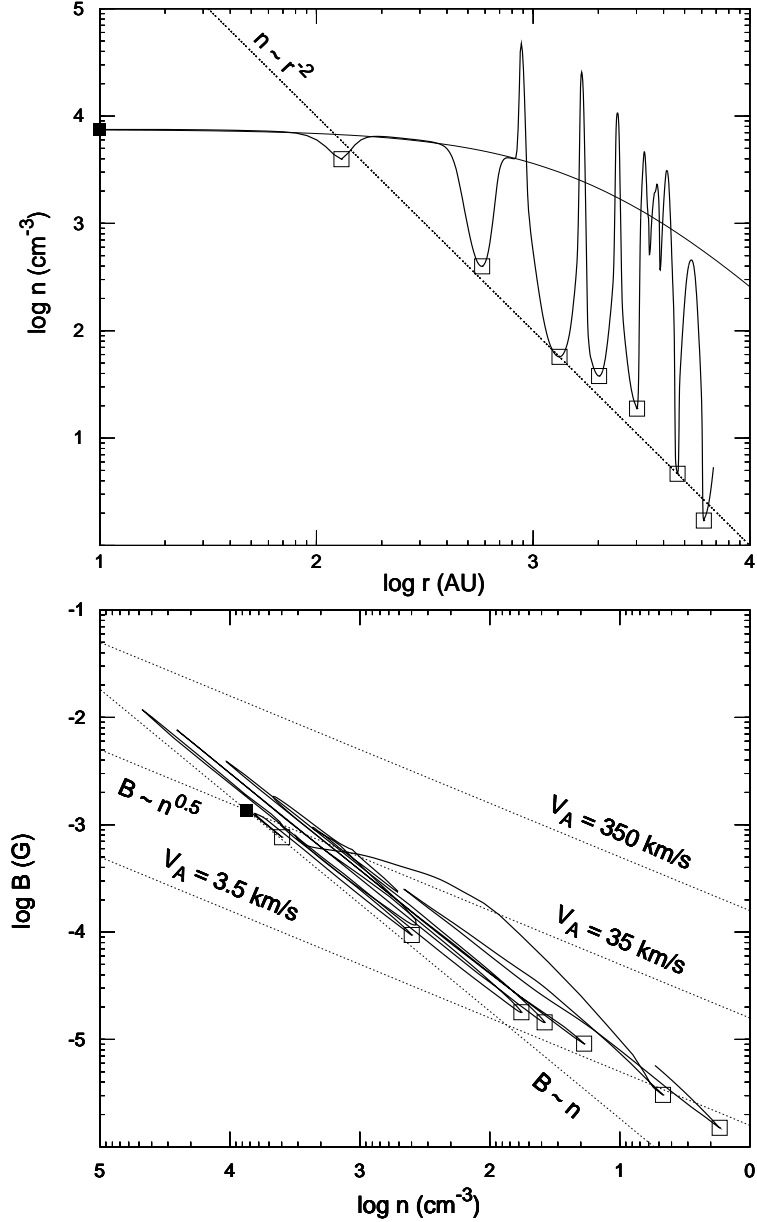}}
\caption{ Top: A snapshot of the density vs.~distance along the axis
of an expanding, variable-velocity magnetized jet, taken once the first bow shock has
left the grid to the right. The sharp peaks and valleys are shocks and
rarefactions, respectively, that form as the flow evolves. Once strong rarefactions
form they follow an approximate n $\sim$ r$^{-2}$ law. The solid curve is the steady-state solution.
Bottom: Same as top but for the magnetic field plotted vs.~density. Shock waves move
the curve to the upper left, and rarefactions drop it to the lower right. The
locus of points along the flow follows an approximate power law, B $\sim$ n$^p$,
with p $\sim$ 0.85.  The simulation begins at the filled-in square, and the strongest
rarefactions are denoted by open squares in both plots.}
\end{figure}
\vfill\eject

\begin{figure}
\vbox to 7.0in{\includegraphics{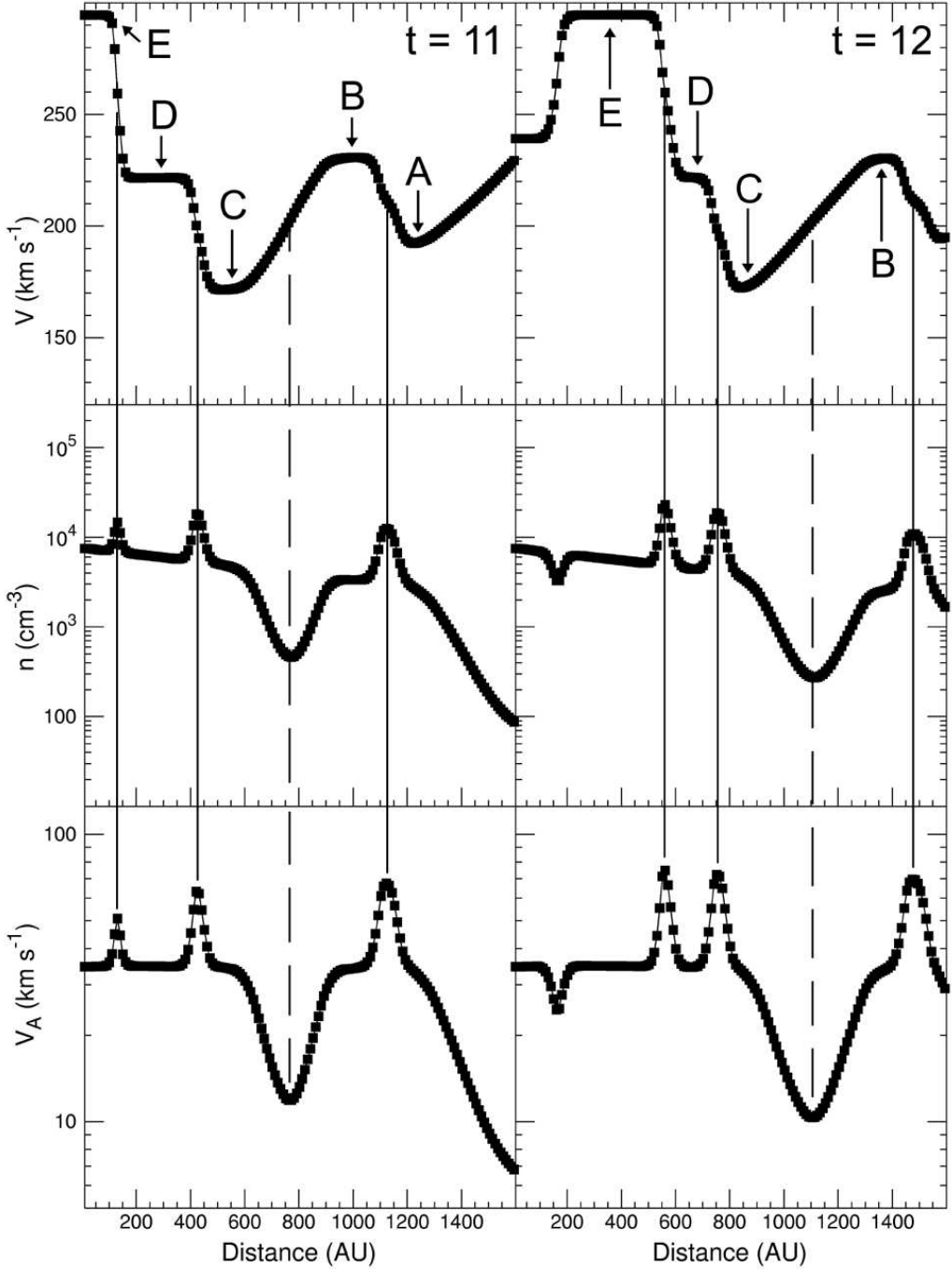}}
\caption{A typical sequence of velocity perturbations, labeled A $-$ E. The top,
middle, and bottom panels are the velocity, density, and Alfven speed, respectively.
Areas of compression are marked by solid vertical lines, and strong rarefactions by
vertical dashed lines. 
The left and right panels show the first 1600 AU of the simulation at times that
correspond to 11 and 12 input velocity pulses, respectively. The leading bow shock
is located well to the right of the figures.
In this, and subsequent figures the plots depict conditions along the axis of
the jet.  Further parameters of the simulation are discussed in the text.
}
\end{figure}
\vfill\eject

\begin{figure}
\vbox to 7.5in{\includegraphics{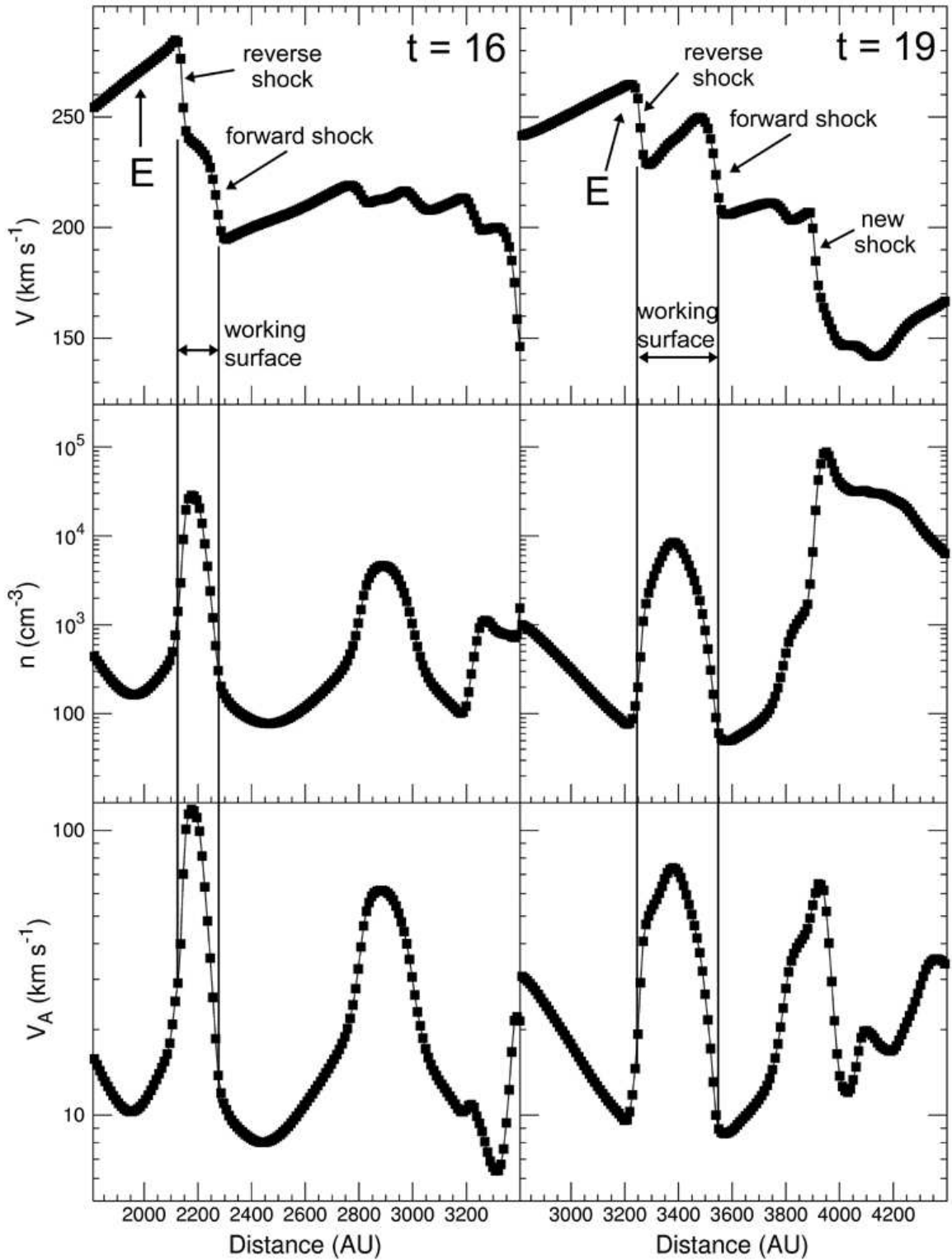}}
\caption{Same as Fig.~2 but at two later times. The evolution of the working surface
of perturbation E is discussed in the text.}
\end{figure}
\vfill\eject

\begin{figure}
\vbox to 3.0in{\includegraphics{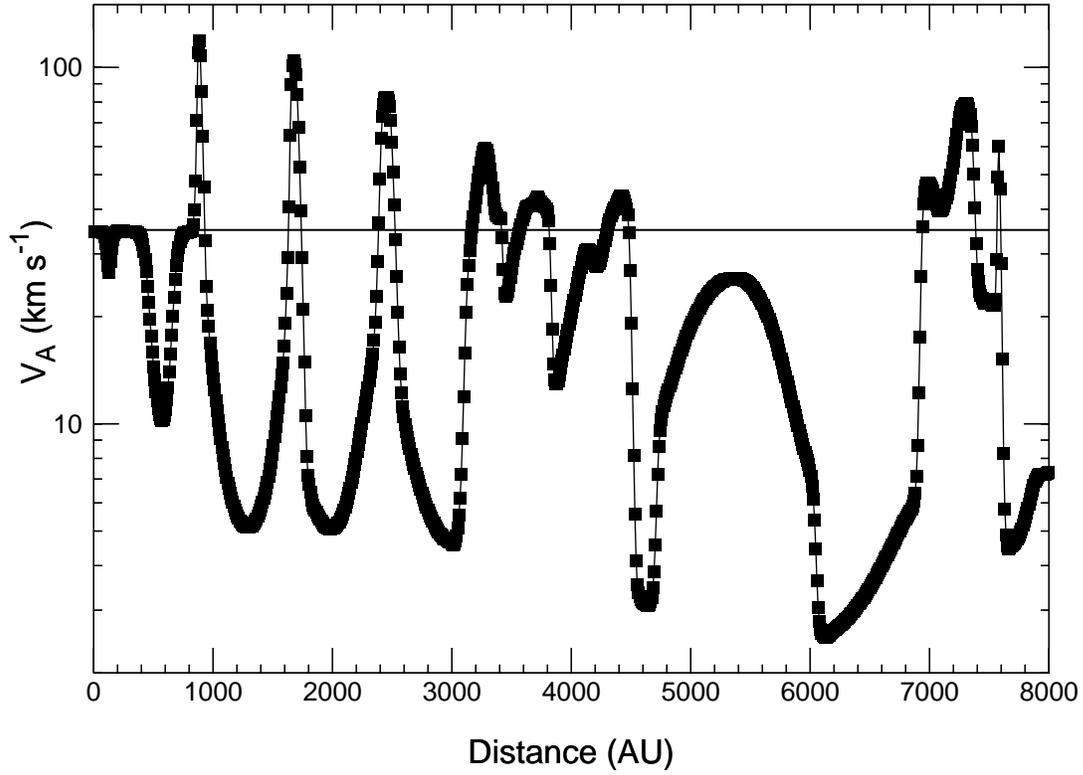}}
\caption{A plot of the Alfven speed at the end of the simulation along the
axis of the jet. The leading bow shock has propagated off the end of the grid.
Magnetic flux concentrates into a few dense knots. Most points fall below the
steady-state solution depicted as a solid line at 35 km$\,$s$^{-1}$.}
\end{figure}

\begin{figure}
\vbox to 6.0in{\includegraphics{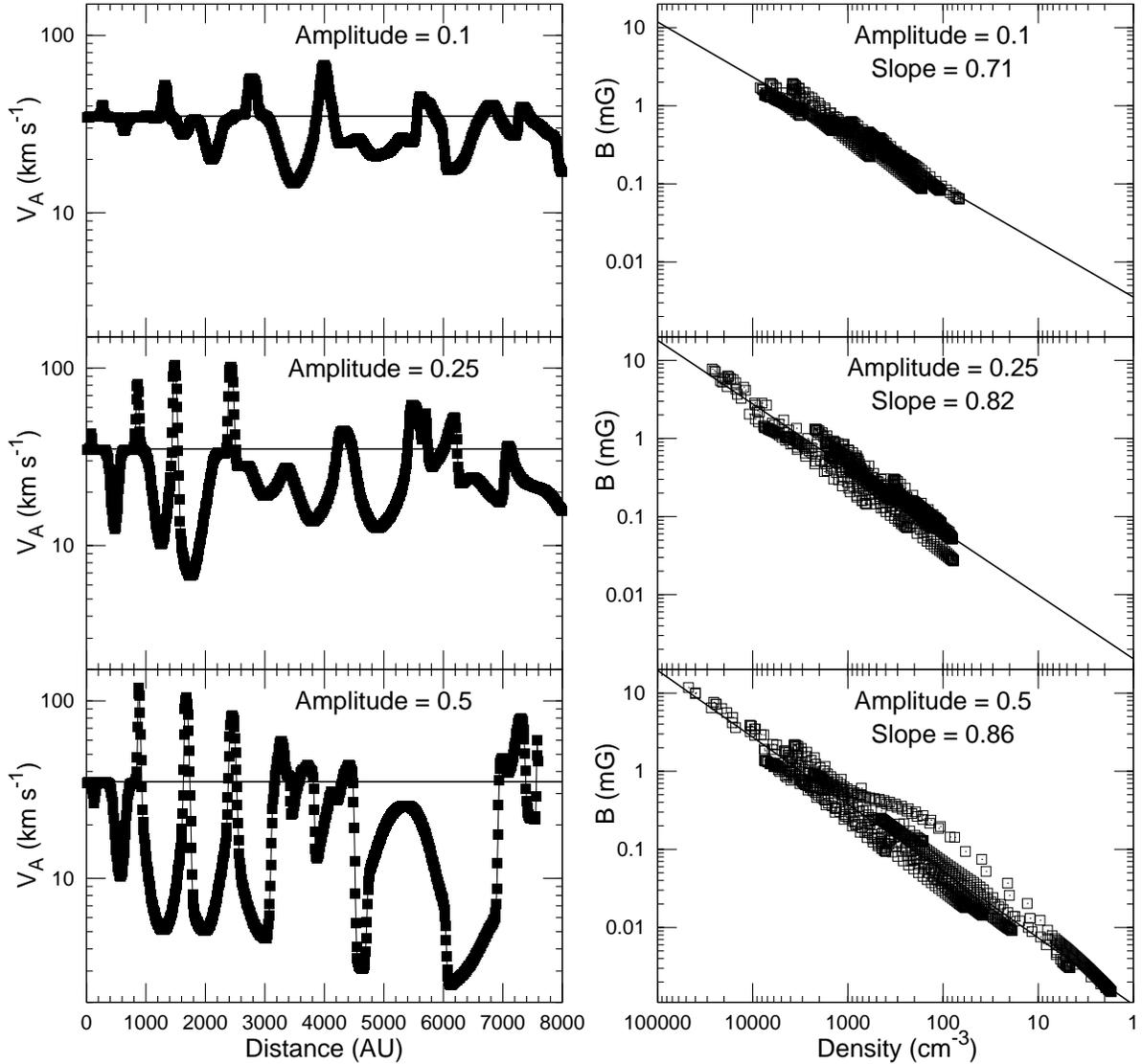}}
\caption{Left: Plots of the Alfven speed V$_A$ \hbox{vs.} distance 
for three different maximum perturbation amplitudes. The horizontal line marks 
V$_A$ = 35 km$\,$s$^{-1}$, which remains constant with distance
when the input flow velocity does not vary. Lower-amplitude simulations have more
modest compressions and rarefactions, but the effect of the perturbations in all
cases is to concentrate high areas of V$_A$ into a few cells, while a typical
value of V$_A$ declines with distance.
Right: Analagous plots of the magnetic field B \hbox{vs.} density n show that 
B $\sim$ n$^p$, with 0.5 $<$ p $<$ 1.
Higher amplitude perturbations produce correspondingly larger 
changes in both B and n.
}
\end{figure}
\vfill\eject

\normalsize

\end{document}